# How students blend conceptual and formal mathematical reasoning in solving physics problems

Eric Kuo
Department of Physics
University of Maryland, College Park

Michael M. Hull
Department of Physics
University of Maryland, College Park

Ayush Gupta
Department of Physics
University of Maryland, College Park

Andrew Elby
Department of Teaching and Learning, Policy and Leadership
University of Maryland, College Park

Abstract:

Current conceptions of quantitative problem-solving expertise in physics incorporate conceptual reasoning in two ways: for selecting relevant equations (before manipulating them), and for checking whether a given quantitative solution is reasonable (after manipulating the equations). We make the case that problem-solving expertise should include opportunistically blending conceptual and formal mathematical reasoning *even while* manipulating equations. We present analysis of interviews with two students, Alex and Pat. Interviewed students were asked to explain a particular equation and solve a problem using that equation. Alex used and described the equation as a computational tool. By contrast, Pat found a shortcut to solve the problem. His shortcut blended mathematical operations with conceptual reasoning about physical processes, reflecting a view – expressed earlier in his explanation of the equation – that equations can express an overarching conceptual meaning. Using case studies of Alex and Pat, we argue that this opportunistic blending of conceptual and formal mathematical reasoning (*i*) is a part of problem-solving expertise, (*ii*) can be described in terms of cognitive elements called *symbolic forms* (Sherin, 2001), and (*iii*) is a feasible instructional target.

# Introduction

The science education literature on quantitative problem solving emphasizes the importance of incorporating conceptual reasoning in two phases of problem solving: (1) initial qualitative analysis of the problem situation to determine the relevant mathematical equations and (2) interpretation of the final mathematical answer, to check for physical meaning and plausibility (P. Heller, Keith, & Anderson, 1992; Redish & Smith, 2008; Reif, 2008). Without disputing the importance of these phases of problem solving, we note that almost no research has focused on the "mathematical processing" stage where the equations are used to obtain a solution. In this paper, we investigate different ways that students can process equations while problem solving. We argue that a feature of problem solving expertise — and a feasible instructional target in physics, chemistry, and engineering courses — is *blended processing*, the opportunistic *blending* of formal mathematical and conceptual reasoning (Fauconnier & Turner, 2003; Sherin, 2001) *during* the mathematical *processing* stage. In other words, we argue that expert problem-solving involves exploiting opportunities to use conceptual reasoning in order to facilitate the manipulation of equations themselves.

To make our case, we first review how physics education researchers have conceptualized and taught quantitative problem solving. We then discuss research suggesting the importance of blending conceptual reasoning with symbolic manipulations in quantitative problem solving, and we propose *symbolic forms* (Sherin, 2001) as cognitive resources that facilitate such *blended processing*. Then we use contrasting case studies of two students solving a physics problem, to illustrate what we mean by blended processing. Alex solves the problem by representing the physical situation with a diagram, identifying the relevant physics equations, using those equations to compute a numerical answer, and reflecting upon that answer — in accord with problem-solving procedures taught in physics classrooms (e.g. Giancoli, 2008; Young & Freedman, 2003) and advocated in education research (P. Heller et al., 1992; Huffman, 1997; Reif, 2008; Van Heuvelen, 1991a). Pat, by contrast, blends symbolic equations with conceptual reasoning about physical processes to find a "shortcut" solution. After analyzing Alex's and Pat's responses in detail, we show that some other introductory physics students in our data corpus also do the type of blended processing done by Pat. In documenting what such blended processing can look like for undergraduate students in an introductory physics course, we make the case that (1) the opportunistic use of blended processing is part of quantitative problem-solving expertise; (2) a theoretical construct called *symbolic forms* (Sherin, 2001) contributes to a good cognitive account of Pat's (and the Pat-like students') blended processing; and (3) such blended processing is a feasible instructional target in science and engineering courses.

## Literature review: Conceptualizations of expert problem solving

In this section, we present a common conceptualization of expertise in quantitative physics problem-solving, as well as challenges to a particular aspect of that conceptualization. We limit our discussion to *quantitative* problem solving, because our argument specifically concerns the processing of equations in problem solving.

**Research on expert problem solving and resulting instructional strategies emphasize an initial conceptual reasoning phase**

As a central feature of their professional practice, scientists apply domain-specific knowledge to solve quantitative problems (Redish & Smith, 2008; Reif, 2008; Reif & Heller, 1982). Partly for this reason, developing problem-solving expertise in students has become a central concern of science education researchers and practitioners (Hsu, Brewe, Foster, & Harper, 2004; Maloney, 1994, 2011).

Early research on physics problem solving suggests a difference between experts and novices. Experts tend to start with a conceptual analysis of the physical scenario, which then leads into the mathematics. By contrast, novices tend to start by selecting and manipulating equations that include relevant known and unknown quantities (Larkin, McDermott, Simon, & Simon, 1980; Simon & Simon, 1978). Specifically, on standard textbook physics problems, experts cue into relevant physics principles whereas novices cue into surface features and their related equations (Chi, Feltovich, & Glaser, 1981). Building on these findings, subsequent research has explored the benefits of helping students analyze the problem situation conceptually (Dufresne, Gerace, Hardiman, & Mestre, 1992; Larkin & Reif, 1979) and has incorporated initial conceptual thinking into models of effective quantitative problem solving (J. I. Heller & Reif, 1984; Reif & Heller, 1982).

This research on expert-novice differences has also influenced researchers' formulation of multi-step problem-solving procedures intended for students to learn and apply (P. Heller et al., 1992; Huffman, 1997; Reif, 2008; Van Heuvelen, 1991a, 1991b). These procedures generally include versions of the following steps: (1) perform an initial conceptual analysis using relevant physics principles; (2) use this qualitative analysis to generate the relevant mathematical equations; (3) use equations to obtain a mathematical solution in a "mathematical processing" step; and (4) interpret that mathematical solution in terms of the physical scenario. These procedures incorporate the expert-novice findings by encouraging students to reason conceptually before jumping into mathematical manipulations.

In these strategies, the steps are meant to mirror behaviors exhibited by experts while also remaining accessible enough to be instructional targets. The explicit teaching and enforcement (through grading policies) of these problem-solving procedures has increased the quality and frequency of physical representations used in problem solving, as well as the correctness of students' answers, in comparison to traditional instruction (P. Heller et al., 1992; Huffman, 1997; Van Heuvelen, 1991a).

**Studies of quantitative problem solving have not focused on *how* equations are processed to reach solutions**

The studies described above illustrate a common feature of research on students' quantitative problem solving: to the extent these studies focus on equations, they focus on how students *select* equations rather than on how students *use* those equations after their selection. While this focus has produced important findings and implications for instruction (for example, emphasizing initial conceptual reasoning for selecting relevant equations), it has also limited attention to how students process mathematical equations to obtain mathematical solutions.

In some research, the equations are all treated (either explicitly or implicitly) as computational tools, devices to find unknown values from known values through symbolic and numeric manipulation. This is true of the problem-solving procedures described above (P. Heller et al., 1992; Huffman, 1997; Van Heuvelen, 1991a) and of studies about how successful problem solvers use mathematics (e.g. Dhillon, 1998; Taasoobshirazi & Glynn, 2009).

Other more recent studies have not attended to any aspect of how equations are processed. Walsh, Howard & Bowe (2007) focused mainly on how students *selected* relevant equations rather than on how those equations are subsequently used. Some studies have deemphasized the use of math completely and focused only on students' qualitative analysis, both in instructional interventions (e.g. Mualem & Eylon, 2010) and in finding predictors of problem-solving expertise (e.g. Shin, Jonassen, & McGee, 2003).

The first author (Kuo) did a search through *Science Education*, *Journal of Research in Science Teaching*, *Research in Science Education*, *International Journal of Science Education, American Journal of Physics*, *The Journal of Engineering Education*, *Cognition & Instruction*, and *Journal of the Learning Sciences* from January 2000 to March 2012 and *Physical Review Special Topics – Physics Education Research* from July 2005 (its inception) to March 2012. He looked for articles focusing explicitly on problem solving in which the analysis attended to the possibility of processing equations in multiple ways.  First, the titles of all articles were scanned, and abstracts of articles with titles containing terms such as "problem solving" or "equations" were read.   If the abstract described investigations of components of problem-solving expertise, the article itself was read.  This search found no studies that focused upon the mathematical processing step in quantitative problem solving or described alternatives to using equations simply as computational tools.

**Other studies suggest the importance of blending conceptual reasoning with symbolic manipulations.**

We have shown that research on quantitative problem solving has not attended to different ways that mathematical equations can be used to obtain numerical or symbolic solutions. This paper presents two different ways that such equations may be used: (i) as computational tools, manipulated to solve for unknown quantities, or (ii) blended with conceptual meaning to produce solutions (or progress toward solutions). But why is this difference significant?

Other pockets of research suggest that using equations without looking to their conceptual meaning *during the processing* can, in certain situations, reflect a *lack* of expertise. In mathematics education research, for example, Wertheimer (1959) asked students to solve problems of the following type: (815+815+815+815+815)/5 = ?. Students who solved the problem by computing the sum in the numerator and then dividing by 5 missed a possible shortcut around explicit computation: Using the underlying conceptual meanings of addition and division to realize that the solution is 815, without doing any computations. Students who missed the shortcut had demonstrated understanding of the mathematical procedures, but not of the underlying conceptual meaning. Additionally, Arcavi (1994) suggested the importance of *symbol sense*: an ability to reason conceptually about symbols. This *symbol sense* includes the ability to interpret the conceptual meaning behind symbolic relationships, generate expressions from intuitive and conceptual understanding, and decide when and how best to exploit one's conceptual understanding of symbols.

Redish and Smith (2008), writing about expert problem solving in science and engineering, also challenged the view that symbolic manipulation should be *a priori* divorced from conceptual reasoning, saying "…because of the fact that the equations are physical rather than purely mathematical, the processing can be affected by physical interpretations." Just as Wertheimer showed that students' conceptual understanding of *mathematical* operations influenced how they carry out calculations in an arithmetic problem, Redish and Smith suggest that students' interpretations of equations in terms of the physical scenario can influence how they use the equations in solving physics/engineering problems.

Again, we do not dispute the instructional value of prior research on problem solving procedures that emphasize conceptual reasoning at the start and the end of problem-solving. As noted above, an instructional emphasis on such procedures has helped students to produce more and better representations and to produce correct solutions more frequently. However, Wertheimer, Arcavi, and Redish and Smith suggest the importance of focusing on *how* students process equations in their quantitative problem solving, a focus not present in the quantitative problem solving literature. Specifically, these researchers argue that blending conceptual reasoning with mathematical formalism in the processing of equations — what we refer to as *blended processing* — may be productive and reflect greater expertise in some situations than using an equation simply as a computational tool. In the next section, we discuss the use of *symbolic forms*, which we argue is one specific way that blended processing can occur in physics problem solving.

**Symbolic Forms: a blend of conceptual reasoning and mathematical formalism**

In arguing that problem solving does not necessarily proceed from direct application of canonical physics principles, Sherin (2001) proposed the existence of knowledge structures called *symbolic forms,* which link mathematical equations to intuitive conceptual ideas. Specifically, in a *symbolic form*, a *symbol template* is blended with a *conceptual schema*.

A *symbol template* represents the general structure of a mathematical expression without specifying the values or variables. For example, $\Box = \Box$ is the symbol template for Newton's 2nd law ($F = ma$), while the symbol template for the first law of thermodynamics, $\Delta E = Q + W$, is $\Box = \Box + \Box$. Each symbol template is not unique to a single equation. For instance, the symbol template $\Box + \Box + \Box$ can describe both the expression $x_0 + v_0 t + 1/2\ at^2$ and the expression $P_0 + 1/2\ \rho v^2 + \rho gh$.

A *conceptual schema* is an informal idea or meaning that *can* be (but does not *have* to be) represented in a mathematical equation or expression. One example is the idea that *a whole consists of many parts*. For example, an automobile can be seen as an assembled whole of many parts such as the engine, the transmission, and the chassis; a wedding guest list can be conceptualized as consisting of the close relatives, the close friends, business contacts, and others; an essay might be viewed as the compilation of the introduction, the main body of argument, and the conclusion. Similarly, a physics student's conceptual understanding of the total mechanical energy of a system may be grounded in the idea that it is comprised of many different types of energy: kinetic, gravitational potential, spring potential, and so on.

Another conceptual schema, applicable to intuitive reasoning about a game of tug-of-war or about a marriage between a spendthrift and a miser, is the idea of *opposing influences*. In physics, this conceptual schema may apply to a student's conceptual understanding of a falling object, where air resistance opposes the influence of gravity (Sherin, 2001).

As these examples illustrate, a conceptual schema in Sherin's framework is an *intuitive* idea used in everyday, nonscientific reasoning, not a formal scientific concept. A student's understanding of a formal scientific concept (such as mechanical energy) can draw upon these intuitive conceptual schemata (such as *whole consists of many parts*); but the conceptual schema also plays a role in students' reasoning about other subjects, such as wedding guest lists.

A *symbolic form* is a cognitive element that blends a symbol template with a conceptual schema, such that the equation is interpreted as expressing meaning corresponding to the conceptual schema. For example, the *parts-of-a-whole* symbolic form blends the symbol template "$\Box = \Box + \Box + \Box + ...$" with the conceptual schema of *a whole consisting of many parts*. The box on the left side of the equation takes on the meaning of "whole" and the boxes on the right side take on the meaning of "parts." A student who uses the *parts-of-a-whole symbolic form* to interpret the equation $E = \frac{1}{2}mv^2 + mgh + \frac{1}{2}kx^2$ would say that the overall (whole) energy of the system consists of the sum of three separate parts (kinetic, gravitational potential, and spring potential). So, when a symbolic form is used, the reasoning is neither purely formal mathematical nor purely conceptual; it is blended into a unified way of thinking that leverages *both* intuitive conceptual reasoning and mathematical formalism. By contrast, a writer thinking about her essay might think of the parts of her essay (Introduction, etc.) using the conceptual schema corresponding to *parts-of-a-whole*, but is unlikely to think of adding those parts in an equation. Only when a conceptual schema is blended with an equation's symbol template is a symbolic form present in the person's reasoning.

Sherin (2001) observed students using symbolic forms in two ways. One was to produce novel equations from an intuitive conceptual understanding of a physical situation. For example, figuring out how much rain would hit them in a rainstorm, a pair of students wrote the equation [*total rain*] = [*#raindrops/s*] + *C*. Their explanation of this equation reflected the use of *parts-of-a-whole*. Specifically, they said the total rain would come from two sources: the amount falling on top of the person, indicated by [*#raindrops/s*], and the amount striking the front of the person as they walk forward, indicated by *C*. Other research has also supported the explanatory power of symbolic forms in models of how students translate physical understandings into mathematical equations (Hestenes, 2010; Izsák, 2004; Tuminaro & Redish, 2007).

The other way in which Sherin's subjects used symbolic forms was to interpret mathematical equations in terms of a physical scenario, using functional relations expressed by the equation. For example, after deriving the terminal velocity of a falling object, $v = mg/k$, several students noticed that the mass, $m$, was in the numerator. Students interpreted this as meaning that a heavier object reaches a greater terminal velocity. Sherin modeled these students as using the *prop+* (positive proportionality) symbolic form — a blend of the symbol template [...x... / ….....] with the conceptual schema that one quantity increases as another one increases— to read out a physical dependence from the mathematical equation. Later, Sherin (2006) hypothesized that *prop+* was tied to physical notions of effort and agency, what we see as "cause-and-effect." Other researchers have also used symbolic forms to model how students translate from mathematical solutions into physical understanding (Hestenes, 2010; Tuminaro & Redish, 2007; VanLehn & van de Sande, 2009).

These two ways in which Sherin (2001) saw students use symbolic forms correspond roughly to the two "steps" involving conceptual reasoning, as described by the problem solving literature: (i) translating conceptual understanding of a physical scenario into mathematical equation(s) at the start of problem solving and (ii) giving a physical interpretation of a mathematical solution at the end[1]. We have seen no studies that look at how symbolic forms-based reasoning — or blended conceptual and formal mathematical reasoning more generally — might enter into the "mathematical processing" step in quantitative problem-solving.

**Building on the literature to explore the mathematical processing step**

We contribute to the literature on quantitative problem solving in three ways. First, we focus on different ways in which mathematical equations can be used to reach a solution. This is a relatively unexplored topic, as discussed above. We find that equations can be used as tools for symbolic or numerical manipulations or as tools in blended processing. As we argue, the opportunistic use of such blended processing can reflect greater expertise than symbolic or numerical computation (Arcavi, 1994; Redish & Smith, 2008; Wertheimer, 1959).

Second, we argue that symbolic forms help to explain the patterns we document below in students' problem solving and are therefore productive analytical tools for researchers trying to understand the nature of blended processing.

Third, we argue that symbolic forms are also plausible targets for instruction in introductory physics, partly because they rely on intuitive rather than formal (or discipline-specific) conceptual schemata. Students can therefore productively use symbolic forms *while* they are still learning difficult physics concepts.

## Methods and Data Collection

### Interview Context

Our data set consists of videotaped interviews with 13 students enrolled in a first-semester, calculus-based, introductory physics course at a large, public university in the United States. The course, geared toward engineering majors, covers mechanics. These students were interviewed between fall 2008 and spring 2011, and our recruitment did not target any particular demographic. The interviews lasted about one hour.

The two subjects on whom we focused our analysis, "Alex" and "Pat" (pseudonyms), were interviewed one and a half months into the course in fall 2008. By that time, the course had covered kinematics, including objects falling under the influence of gravity, which is the topic of the interview prompts discussed below. We chose Alex and Pat for fine-grained analysis because they were among the first students we interviewed and because the strong differences between their responses motivated us to seek an explanation for those differences.

**Interview protocols**

We designed the semi-structured interviews to probe engineering majors' approaches to using equations while solving quantitative physics problems. Specifically, we wanted to explore what formal mathematical and conceptual tools they bring to bear and which epistemological stances they take toward the knowledge they use. Moreover, we designed some prompts specifically to probe whether and how students use blended processing when the opportunity arises. To that end, we had students think aloud while solving specific problems. We also asked them to explain the meaning of both familiar and unfamiliar equations and to discuss more generally how they know when they "understand" an equation. The complete protocol is available online (http://hdl.handle.net/1903/12947). Our analysis in this paper focuses on the first two prompts in the interviews.

*Prompt 1: Explain the velocity equation*

The interviewer shows the student the equation $v = v_0 + at$ and asks, *Here's an equation you've probably seen in physics class. How would you explain this equation to a friend from class?*

*Prompt 2: Two Balls Problem*

*(a) Suppose you are standing with two tennis balls on the balcony of a fourth floor apartment. You throw one ball down with an initial speed of 2 meters per second; at the same moment, you just let go of the other ball, i.e., just let it fall. I would like you to think aloud while figuring out what is the difference in the speed of the two balls after 5 seconds – is it less than, more than, or equal to 2 meters per second? (Acceleration due to gravity is 10 m/s$^2$.)* [If the student brings it up, the interviewer says to neglect air resistance][2]

(b) [Only if student solved part (a) by doing numerical calculations] *Could you have solved that without explicitly calculating the final values?*

In designing the Two Balls Problem, we anticipated that some students would find something like the following shortcut: according to the velocity equation, $v = v_0 + at$, since both balls *gain* the same amount of speed over 5 seconds (with the gain given by the mathematical term *at*), the final difference in speeds equals the initial difference in speeds, in this case 2 meters per second. Note that the shortcut uses blended processing, interpreting the equation conceptually as "the final velocity is the initial velocity plus the change in velocity." We wanted to see if students would notice this shortcut solution or something like it, either on their own or in response to the follow-up prompt about whether the problem could have been solved without numerical calculations.

Although this problem is not as complex or difficult as some textbook introductory physics problems, it is still a problem that we would want students to be able to solve; reform-oriented physics classes often ask questions such as this (e.g. Redish & Hammer, 2009). Moreover, we believe that this problem's conceptual shortcut afforded a way to investigate differences in how students process mathematical equations in problem solving; we suspected that blended processing would contribute to finding the shortcut. Although blended processing is not explicitly sought in most textbook physics problems, the *opportunistic* use of such reasoning when it is possible is an aspect of problem-solving expertise, as argued above. In asking the Two Balls Problem, we hoped to elicit data on whether and how students perform blended processing.

**Analysis phase 1:  Alex and Pat**

We began with fine-grained, qualitative analysis of Alex's and Pat's approaches to the Two Balls Problem and their explanations of the velocity equation, $v = v_0 + at$. Our goal was to characterize how the two subjects were conceptualizing the equation and its role in problem solving.

*Phase 1a:  Two Balls Problem (prompt 2)*

To start, we looked at video and corresponding transcript of the two subjects' responses to the Two Balls Problem (prompt 2). In trying to characterize how they were thinking about and using the velocity equation, we analyzed the solutions they were speaking and writing while thinking aloud. However, other markers of their thinking, such as word choice, pauses, and speech rhythm also informed our interpretations.

Although these markers are discussed by the discourse and framing analysis literature,
(Gee, 1999; Tannen, 1993), we do not claim to be doing discourse or framing analysis.

After formulating possible characterizations, we performed a line-by-line analysis
through this particular section of the transcript (response to prompt 2) for confirmatory
and/or disconfirmatory evidence. In this way we refined and narrowed down the plausible
characterizations of how Alex and Pat were thinking about the equation (Miles &
Huberman, 1994).

*Phase 1b: The velocity equation (prompt 1)*

With Alex and Pat, we had reached tentative consensus in phase 1a about how
they were thinking about and using the velocity equation in solving the Two Balls
Problem (though with Pat, we lacked evidence to decide between two subtly different
interpretations). We then analyzed their responses to prompt 1, which asks how they
would explain the velocity equation to a friend, in order to look for confirmatory or
disconfirmatory evidence for our phase 1a interpretation. Using the same analytical tools
as in phase 1a, we tried to characterize how they were thinking about the equation in the
context of explaining it. We then compared our characterizations to what we had found in
phase 1a. As discussed below, the alignment was strong. By providing detailed analysis
below, and also the complete transcript (http://hdl.handle.net/1903/12947), we give
readers the opportunity to check if we matched up our characterizations in phase 1a and
1b when it was not warranted.

**Analysis phase 2:  Looking at other students**

Although the literature has not documented Pat-like blended processing of equations, our experiences as physics instructors and our subsequent interviews led us to think his reasoning might not be idiosyncratic. Furthermore, while the presence or absence of blended processing strongly distinguishes Pat's reasoning from Alex's, other students might not fall so cleanly on one side of this distinction or the other.

To explore these issues, we analyzed the responses of the 11 other students we interviewed  (besides Alex and Pat). Specifically, the first and second author independently coded whether each student (1) used blended processing to find a shortcut solution to the Two Balls Problem, either initially or in response to our follow-up prompt asking if the problem could be solved without plugging in numbers; and (2) gave an explanation of the velocity equation that combined the symbol template with a conceptual schema — i.e., a symbolic forms-based explanation. The two independent coders initially agreed on 9 of 11 codes for code 1 and on all 11 codes for code 2.

Since the number of students in this study is small, we are not aiming to make statistically significant claims about patterns of reasoning.  We use these data instead to bolster our arguments that blended processing is a useful analytical lens for understanding students' reasoning and a feasible instructional target.

# Results of analysis phase 1a: The Two Balls Problem

**How Alex used the velocity equation while solving the Two Balls Problem**

*Alex solved the problem with a numerical calculation*

Alex started by drawing a diagram of the two balls and labeling their speeds (figure 1). After deciding to use the velocity equation to solve this problem, Alex paused and remarked that she did not have a value for *a* for the equation, $v = v_0 + at$. She realized that *a* should be 9.8 and wrote this value in her diagram. The interviewer interjected and said that she could use 10 if she wanted, and Alex responded that using 10 is probably easier. She then explicitly solved for the velocities of the thrown and dropped balls after five seconds and wrote down the difference. Figure 1 shows all of her work[3].

Figure 1: Alex's written work on the Two Balls Problem

After working out the speeds of the two balls to be 50 m/s and 52 m/s, Alex explained her thought process:

A31 Alex: …Ok, so after I plug this into the velocity equation, I use the acceleration and the initial velocity that's given, multiply the acceleration by the time that we're looking at, five seconds, and then once I know the velocities after five seconds of each of them, I subtract one from the other and get two. So the question asks "is it more than, less than, or equal to two?" so I would say equal to two.

("A31" refers to the 31st conversational turn in the interview with Alex.)

In this segment, Alex followed a set of steps similar to the ones advocated in research-based problem-solving strategies: draw a picture of the situation (which can include labeling known values), choose the equation relevant to the physical situation, and calculate the desired unknowns to answer the question. Alex executed this procedure smoothly, with the only pause coming when she considered the value of the acceleration. We tentatively conclude that, during this segment, Alex used the velocity equation as a tool for numerical computation.

*Alex exhibited hints of conceptual reasoning, but not stably integrated with the velocity equation*

After Alex gave her solution to the Two Balls Problem, the interviewer asked if someone could have answered the question without explicitly solving for the velocities of both balls. We designed this follow-up prompt to get at whether students (i) might have implicitly used blended processing in a way that their think-aloud and written solution did not reveal, or (ii) might get cued into blended processing by this prompt.

Alex's first answer was yes, but when asked to elaborate, she seemed unsure:

A35 Alex: Well, I'd have to think about it, since you're dropping one and throwing one. If you're, I mean I guess if you think you're throwing one 2 m/s and the other has 0 velocity since you're just dropping it, its only accelerating due to gravity, you can just say that since you know one is going at 2 m/s, it's going to get there 2 m/s faster, so 5 seconds faster, it would get there 2 seconds… er… it's going 2 m/s faster, I guess.

A36 Interviewer: OK. So, they would say that you threw one, so this was getting 2 m/s faster. So what happens 5 seconds later?

A37 Alex: Uh…it's going…uh…I don't know. (laughs)

As the interviewer followed up, Alex continued to sound less and less sure about her answer. Finally, she changed her mind:

A51 Interviewer: So you're saying that they need not have actually plugged in the numbers? Is that what I'm hearing?

A52 Alex: No, I think you'd have to plug in the numbers because…uh…I mean you just would to be sure. I guess you, I don't think you can just guess about it.

In line A35, Alex attempted to show how the Two Balls Problem can be answered without an explicit calculation. One interpretation of this exchange is that she never came up with a firm conceptual explanation for how to solve the problem without calculations, as evidenced by her mixing up the units ("...so 5 seconds faster, it would get there 2 seconds, err, it's going 2 meters per second faster..."). A different interpretation is that she was trying to express the following conceptual argument: since both balls are accelerating due to gravity only, both balls will gain the same amount of speed, so the thrown ball will be traveling "2 meters per second faster" (line A35). Either way, she backed off this line of reasoning in lines A37 and A51, possibly because she felt on the spot trying to answer the interviewer's questions (evidenced by her greater hesitancy than when she presented her original solution). Our point is that, either way, any conceptual reasoning in line A35 was not *stably* integrated with Alex's mathematical, symbolic reasoning. Evidence for this lack of stable integration comes from (*i*) the lack of explicit mention of the equation or implicit reliance on its structure in line A35, and (*ii*) her view in line A51 of the calculation as a way to "be sure" of non-calculation-based reasoning, which is more of a "guess" than something reliably connected to the calculation in some way.

In summary, the procedural way in which Alex solved the problem, along with the lack of a stable connection between the equation and conceptual reasoning evidenced by her follow-up comments, points us toward the conclusion that Alex, in this context, was viewing the velocity equation as a computational tool for calculating a final velocity given an initial velocity, an acceleration, and a time. We will put this initial interpretation to the test below when we analyze how Alex explained the velocity equation. But first, to emphasize the contrast between Alex and Pat, we present Pat's solution to the Two Balls Problem, which he solved without calculating the final velocity of either ball.

**How Pat used the velocity equation while solving the Two Balls Problem**

*Pat solved the problem without plugging in numbers*

Like Alex, Pat also turned to the velocity equation. However, he used it very differently from the way Alex did:

> P41  Pat: …Well, the first thing I would think of is the equations. The velocity, I suppose, is the same equation as that other one [the velocity equation he had just explained in prompt 1], and I'm trying to think of calculus as well and what the differences do. So the acceleration is a constant and that means that velocity is linearly related to time and they're both at the same…so the first difference is the same. I *think* it's equal to two meters per second.

Later, when asked by the interviewer how he got this answer, Pat elaborated on his solution a little more:

> P45  Pat: So the first differences are the same.
>
> P46  Interviewer: Mhm.
>
> P47  Pat: And if the first differences are the same then the initial difference between the two speeds should not change.

When asked, Pat explained that the term "first differences" comes from his high school algebra class, where sets of data points would be analyzed by taking "delta y over delta x," which is called the "first difference." So, "first difference" connects at least roughly to the notion of slope.

A few moments later, Pat stated that "there's a couple of methods of attacking" the problem if he gets stuck. Pat then further discussed different ways to solve the Two Balls Problem:

> P61   Pat: So if I started from thinking about the equations and I'm not quite sure whether the velocities are changing at the same rate, then like sometimes I'll use several [solution methods] and see if they're consistent. Then I could switch to thinking about the derivatives of the velocity and I'll think, ok, so the initial conditions are off by 2 and then the velocities are changing at the same rate so that should mean they stay at 2…

Pat did not follow a set of steps similar Alex's. Instead, his solution is a shortcut around an explicit calculation: since the velocities of the two balls change at the same rate, the difference between those two velocities stays the same. Notably, the velocity equation ($v = v_0 + at$) plays a role in his shortcut, but his reasoning is not purely symbolic. Pat started his solution (line P41) by referring to the velocity equation, but he used it to point out that "velocity is linearly related to time" which led him to say that the "first difference" is the same. Since "first difference" is similar to the idea of slope, this aligns with his reasoning in line P61, where he offered a similar argument in terms of derivatives and explicitly stated that the velocities are changing at the same rate.

We use these case studies of Alex's and Pat's reasoning to make two points. First, although both Alex and Pat reasoned productively and correctly solved the Two

Balls Problem, Pat's style of reasoning is not currently described in the problem solving literature. Illustrating this productive reasoning can help expand our understanding of how students approach similar problems.

Second, we view Pat's reasoning as aligning better with expert problem solving than Alex's more procedural approach does. Pat saw multiple solution paths, which he related to one another, while Alex saw just one. Pat flexibly used the available information, which is a component of what Hatano & Inagaki (1986) call "adaptive expertise," while Alex's approach appeared more step-by-step. Also, Pat connected *conceptual meaning* to mathematical formalism: the idea that if two things change at the same rate, then the difference between them stays the same. As Wertheimer (1959), Arcavi (1994), and Redish & Smith (2008) argue, such blended processing indicates a deeper, more expert understanding than simply using the formalism. Other researchers also emphasize the deeply connected nature of expert knowledge (Chi et al., 1981; Chi, Glaser, & Rees, 1982; Reif, 2008; Reif & Heller, 1982), though they do not explicitly discuss connections between informal conceptual knowledge and mathematical formalism. Linking conceptual reasoning to mathematical formalism as Pat does — using blended processing of the velocity equation — is arguably an example of forging or exploiting such connections. These connections can support quick and intuitive solutions, as well the flexible coordination of multiple strands of reasoning, as Pat illustrates.

Although we have hypothesized that Pat was reasoning by connecting a mathematical equation to an intuitive conceptual schema (i.e., "if two things change by the same amount, the difference between them stays the same"), we see at least one plausible alternative account. It is possible that Pat's reasoning was driven by a formal rule of mathematical operations and objects, such as "if the derivative/first difference of two quantities is equal, then the difference between them doesn't change." Our phase 1b analysis of Pat's explanation of the velocity equation will help distinguish between these possibilities, in the end favoring our initial interpretation that Pat was blending conceptual and symbolic reasoning.

## Results of analysis phase 1b: Explaining the velocity equation

So far, we have tentatively concluded that in the context of the Two Balls Problem, Alex viewed the equation as a computational tool while Pat was more flexible, reasoning with the equation to find a conceptual shortcut. To find confirmatory or disconfirmatory evidence, we now look to Alex's and Pat's responses to prompt 1, which asked how they would explain the velocity equation to a friend. We show how the absence of symbolic forms-based reasoning in Alex's explanation and the presence of such reasoning in Pat's explanation helps us understand the differences in their approaches to the Two Balls Problem.

### Alex explains the velocity equation as a computational tool

Alex initially seemed puzzled by this question but eventually answered:

A10   Alex:  Umm…Ok…well…umm…I guess, first of all, well, it's the equation
        for velocity.  Umm…well, I would…I would tell them that it's uh… I mean,
        it's the integral of acceleration, the derivative of {*furrows brow*} position,

right? So, that's how they could figure it out…I don't know. I don't really {*laughs*}, I'm not too sure what else I would say about it. You can find the velocity. Like, I guess it's interesting because you can find the velocity at any time if you have the initial velocity, the acceleration, and time…

Alex's explanation here has two main parts. First, the velocity equation is defined through its relation to other kinematic equations; it is the integral of acceleration and the derivative of the position equation. Second, the equation can be used as a computational tool: to calculate the velocity at some time if you know the other values in the equation.

The interviewer then asked if that is what she would have said on an exam. She said "no" and elaborated:

A14  Alex: Um…well, it depends on what it was asking…'cause I feel like your question's kind of vague, but, I mean, I would probably just say 'it's the velocity equation' {*nods and laughs*}. I mean, if it was a more specific question, I could probably like, elaborate, I guess.

Finally, Alex was asked to explain the equation to a 12-year old who knows math but does not really know physics.

A16  Alex: Well, these two sums will tell you how fast something is going. If you know how fast it's going when it first starts and after it first starts moving and you know its speed when it first starts moving, and you know a certain point in time. You're looking at a certain point in time at which the object is moving, and you know how fast it's changing its speed, you can find how fast it's moving at that time, or you can find out the acceleration from it if you know how fast it's going at that time.

Unlike her response to a friend from class or on a test, Alex now explicitly described the mathematical variables in terms of physical ideas: "if you know *how fast it's going when it first starts* [$v_0$]...and you know a *certain point in time* [t]... *and you know how fast it's changing its speed* [a]…*you can find out… how fast it's moving at that time* [v]."

Yet, even in light of this conceptual interpretation of the variables, there is evidence that Alex's explanation of the equation as a whole is still as a computational tool, as in line A10: if you know any three variables, you can solve for the fourth. In line A16, Alex explained that you can solve for *v* if you know $v_0$, *a*, and *t*. Alternatively, she stated, you can solve for *a* if you know *v*. This interpretation is coherent with how she actually used the velocity equation to compute the final velocities in the Two Balls Problem.

Although Alex attached physical meaning to the individual variables in the velocity equation, Alex's explanation does not include a symbolic forms-based interpretation of the velocity equation *as a whole*. There is no evidence of an intuitive conceptual schema deeply associated with the symbol template reflecting the structure of the velocity equation ($\square = \square + \square$). The following analysis of Pat's explanation of the equation will provide a contrasting case to clarify what evidence we use to make claims about the presence of a symbolic form in a student's reasoning.

**Pat connected the equation to a physical process**

When asked to explain the equation to a friend from class, Pat started by looking at the units and meaning of the variables:

P2    Pat: Well, I think the first thing you'd need to go over would be the definitions of each variable and what each one means, and I guess to get the intuition part, I'm not quite sure if I would start with dimensional analysis or try to explain each term before that. Because I mean if you look at it from the unit side, it's clear that acceleration times time is a velocity, but it might be easier if you think about, you start from an initial velocity and then the acceleration for a certain period of time increases that or decreases that velocity.

Pat started with the definitions of each variable, as Alex described in line A16;  however, he then provided some preliminary evidence of interpreting the equation in terms of a symbolic form we now introduce, called *Base + Change* (Sherin, 2001, p.514).  In *Base + Change*, the symbol template $\square = \square + \square$ is linked to the intuitive conceptual schema that the *final amount is the initial amount plus the change in that amount*. For example, a careful consumer might have this symbolic form active in her reasoning while balancing her checkbook.

In line P2, Pat provided glimmers of evidence that he was relying upon a *Base + Change* interpretation of the equation as a whole rather than just interpreting the individual variables. First, he signaled a shift away from discussing the meaning of the individual variables when he referred to getting to "the intuition part." Then, by doing dimensional analysis of the term *at*, he indicated that he is starting to think of *at* not only as a product of two different quantities, but as a single term with the same units as $v_0$, which hints at reasoning at the level of the symbol template $\square = \square + \square$. Finally, he transitioned from talking about the individual terms in the equation into an overall "story" of a physical process that the equation represents: "...*you start from* an initial velocity *and then* the acceleration for a certain period of time *increases or decreases* that velocity" (emphasis ours). Part of the evidence for this transition is a shift in Pat's narrative perspective. Up until this point, Pat had been speaking from the perspective of a person working with the equation: how "you" or "I" would use or explain the equation. At the transition, the "you" shifted to an object or person that starts with an initial velocity and then undergoes a change in that velocity. This shift in perspective suggests a shift in meaning, from his previous ideas about the definition of variables and dimensional analysis to something else.

This "something else" for Pat relied upon the conceptual schema associated with the *Base + Change* symbolic form: *the final amount* (in this case, final velocity) *is the initial amount* (initial velocity) *plus the change to that initial amount* (due to acceleration)." Pat's reliance on this conceptual schema would be solid evidence of symbolic forms-based reasoning except that it is not yet absolutely clear whether Pat was connecting this conceptual schema to the symbol template of the equation.

After Pat discussed how $v_0$ or $a$ can take on positive and negative values, the interviewer asked Pat what he meant by "the intuition part" in line P2:

P9    Interviewer: So right when you started you said something about "well, then from the intuitive side."

P10   Pat: Yeah, the problem is dimensions are just numbers really, or units, and it doesn't really explain what's going on in the motion.

…

P15   Interviewer: Ok, so how would you explain it intuitively?

P16   Pat: I would say that an acceleration is the change in velocity, so you start from the velocity you have in the beginning and you find out how the acceleration affects that velocity. Then that would be the significance of each term.

The first sentence in line P16 reiterates the "story" in line P2.  It is the second sentence, "[t]hen that would be the significance of each term," that provides solid evidence for the presence of the *Base + Change* symbolic form in Pat's reasoning. Whereas before we could not be sure the idea of "final equals initial plus change" was connected to the symbol template, here we interpret Pat's explanation as saying that "the velocity you have in the beginning" and "how the acceleration affects that velocity" correspond to the terms $v_0$ and $at$, respectively. This corresponds to a *Base + Change* symbolic form interpretation, where $v_0$ is the *base* velocity and $at$ is the c*hange* to that base velocity.

This explanation contrasts with Alex's. She attached conceptual meaning only to the individual variables in the equation.  Here, Pat interpreted the *whole equation* as representing a process of starting with a base amount and changing that base by some value to obtain the final amount.

This analysis supports and refines our analysis from phase 1a, where we interpreted Pat's reasoning as blending conceptual reasoning with formal mathematics. In phase 1a, we highlighted an alternative possibility: that Pat's reasoning was driven by a *formal* rule rather than an *intuitive* schema rooted in a conceptual understanding of the physical process. Here, however, we see that Pat viewed the velocity equation as expressing an intuitive conceptual schema connected to his conceptual understanding of the physical process of speeding up, and specifically that his reasoning is rooted in the *Base + Change* symbolic form. Pat's initial solution to the Two Balls Problem, as discussed above, came from thinking of the formal equation (lines P41 and P61) and relies on the conceptual reasoning that since the *initial* difference in velocities is 2 meters per second, and because both balls undergo the same *change* in velocity, the *final* difference in velocities is still 2 meters per second. Although it is unclear whether he was applying *Base + Change* to the individual velocities of each ball or directly to the difference in those velocities, he was connecting the equation — its general structure and its linearity — to the intuitive schema of "final equals initial plus (linear) change." Pat's explicit explanation of the velocity equation with *Base + Change* provides confirmatory evidence of blended processing in his solution to the Two Balls Problem.[4]

### Summary of differences between Pat's and Alex's views of the velocity equation

Looking across the first two prompts in the interviews ("explain the velocity equation" and "solve the Two Balls Problem"), we see a key difference in how Alex and Pat connected their conceptual understanding of a physical situation to an equation. Alex's connection was at the level of individual variables, while Pat additionally saw the equation as-a-whole expressing an intuitive conceptual idea about the physical process: the velocity you start with plus the velocity you gain (or lose) due to acceleration over a certain period of time is the velocity you end up with. We have also argued that this difference between Alex's and Pat's conceptualizations of the equation offers explanatory power for many of the differences in their responses to those two prompts, including Alex's explicit calculation and Pat's blended processing.

We are not claiming that Alex does not "have" the *Base + Change* symbolic form, or symbolic forms in general, in her repertoire of cognitive resources. Indeed, she showed evidence of blending symbolic and conceptual reasoning later in the interview. We are only claiming that, for whatever reasons, the *Base + Change* symbolic form was not tied to the velocity equation in Alex's responses to the first two prompts.

Additionally, our point is not only the specific form of Pat's blended processing, but also that Pat productively blended conceptual reasoning with mathematical formalism *at all* to inform his use of equations during problem solving. Throughout the interview, Pat blended conceptual and formal mathematical reasoning in ways that we sometimes, but not always, identify as reflecting the use of a symbolic form.

## Results of analysis phase 2: Looking at other students

To see if Pat was idiosyncratic and if the presence or absence of blending conceptual and formal mathematical reasoning is a useful distinction more broadly, we coded the remaining 11 interviews. Specifically, we coded whether each student (1) used blended processing to find a shortcut solution to the Two Balls Problem, either initially or in response to our follow-up prompt asking if the problem could be solved without plugging in numbers; and (2) gave an explanation of the velocity equation that combined the symbol template with a conceptual schema — i.e., a symbolic forms-based explanation.

**Other examples of blended processing**

Through this coding, we showed that Pat's blended reasoning is not completely idiosyncratic. Five of the eleven other students showed evidence of either a symbolic forms-based explanation of the velocity equation or a blended processing shortcut on the Two Balls Problem. Since we find few examples of such reasoning in the literature, we now present examples from those five students. First, we present two examples of blended processing on the Two Balls Problem, starting with Meg:

[After Meg answers the Two Balls Problem simply using the velocity equation to compute the solution through symbol manipulation, the interviewer asks her if she was surprised by that answer.]

Meg: I expected it to be two, because I just I remember something, you know, that if the acceleration will be the same, gravitational acceleration is the same, so what's, in my mind I just reason out that, you know, if one has more of the speed than the other, because the change is the same. So then it's still going to be, the difference is still going to be the same. If you're changing both by the same amount, then in the end one is going to have the same amount more [that it does initially] than the other.

Here, Meg used the idea that because the balls are both "changing [speed]...by the same amount," the "difference [in speeds] is still going to be the same" as it was initially.

Next, we present Sam's reasoning:

Sam: There's no force acting on them after they have gone, they, one just has initial speed, so we know that, you know, the acceleration, um, you know, multiplied by, you know, they're both in there for, both um subjected to gravity [laughs] for five seconds, so they will both have the, you know, same accelera...they're exposed to the same acceleration for the same amount of time, which would give, you know, an additional velocity, um, of the same. Um, however, you know, we could have this equation, you know. $v_0$, well one, we have an initial velocity of two, and one, we have an initial velocity of zero, so we know that one already is going to be experiencing, you know, faster, er well it will be going faster for two seconds, just before we even start.

Even though at the end Sam stated that one ball will be "going faster for two seconds," his previous utterances suggest that he meant that the thrown ball will be traveling two meters per second faster at five seconds. His argument, though perhaps harder to follow than Meg's, is substantively similar: Because both balls are "exposed to the same acceleration for the same amount of time," they both get "an additional velocity" that is "the same." Therefore, since for "one [ball], we have an initial velocity of two, and [for the other ball], we have an initial velocity of zero," the difference in speeds remains two. Supporting this blended-processing interpretation of Sam's explanation is his follow-up remark after plugging through the equations:

> Sam: You know, even before doing any math, like I said before, we already know that this is going to be, well it's going to be, not faster, but a quantity of two meters per second more regardless of what the acceleration is or time.

So, in both Meg's and Sam's reasoning, we see similarities to the blended processing described in Pat's solution: since both change by the same amount, the final difference equals the initial difference.

Next, we present two students' explanations of the velocity equation that we view as using the *Base + Change* symbolic form. We start with Meg:

> [Meg uses the example of a falling object to explain the velocity equation.]
>
> Meg: So the final velocity, the velocity that it hits the ground with is related to the initial velocity because the object has an initial velocity and if you think about it, if the object is moving and it's constantly changing velocity…You can start off with the initial velocity and then you multiply the change in velocity with time, how much time it took and that should theoretically give you how much the velocity has changed, correct? So if you, so then if you have the initial velocity and you have how much the velocity changed, um, and you add those two together that should, in theory, give you the final velocity.

Stan: …Acceleration is a rate of change in speed and $t$ is the time. $at$, like the whole thing, is what you changed in a period, so I'll say that $v$ $initial$ is the speed that you already had, plus the speed that you changed is the speed that you have right now.

In these two representative cases, the students interpreted the $at$ term in the equation as the change in the initial velocity, which when added to the initial velocity gives the final velocity – which is typical of how the *Base + Change* symbolic form is applied to the velocity equation (Sherin, 2001, pp.514)

**Is blended processing a more broadly useable distinction?**

On the Two Balls Problem, three students used blended processing, seven did not, and one was ambiguous. In explaining the velocity equation: six students used symbolic forms-based reasoning, and five did not[5].

As mentioned above, each of two researchers independently produced 22 codes (two responses each for 11 students). They agreed on 20 of the codes. One of these disagreements was quickly resolved through discussion. The other disagreement, on one student's response to the Two Balls Problem, was deemed "ambiguous," because the evidence could be used to both support and refute the presence of blended processing. This agreement between independent coders and the small number of uncodable responses (one out of 22) suggest that the presence/absence of blended conceptual and formal mathematical reasoning is a usable distinction for other students in our study, and plausibly, for larger populations as well.

# Discussion:  Revisiting what counts as problem solving expertise

We now argue that Pat's (and other students') productive use of blended processing gives us reason to amend the common view of what constitutes good problem solving in physics, and hence, of what strategies instructors should nurture in their students.

**Most research-based problem-solving strategies do not include blended processing**

As discussed in the literature review, research on expert problem solving and instructional problem-solving procedures for fostering such expertise have focused on the finding that novices jump right into manipulating equations without a physical understanding of the problem. Without disputing the importance of research building on this finding, we note that the problem-solving procedures studied and advocated by researchers and instructors do not consider different ways that equations can be used to solve a problem. These procedures typically conceive of the mathematical processing step as manipulating symbols until you obtain an unknown (Giancoli, 2008; P. Heller et al., 1992; Huffman, 1997; Van Heuvelen, 1991a, 1991b; Young & Freedman, 2003). However, blending symbolic manipulations with conceptual reasoning when possible is a part of problem-solving expertise, because such reasoning is adaptive and flexible and because it leads to quicker, more generalizable solutions.

Our contribution to this line of research is to illustrate in some detail this alternative way of using equations in problem solving. We contrasted students who did and did not incorporate conceptual reasoning into their mathematical processing. Through these qualitative case studies, we made the case that blended processing was demonstrably productive for the students who used it in solving the Two Balls Problem: they reached the solution quickly, without having to do extensive calculations.

Moreover, for the physics education research community, which has emphasized the importance of conceptual understanding (Hestenes, Wells, & Swackhamer, 1992; McDermott, 1991; McDermott & Redish, 1999), the blended processing in students' solutions to the Two Balls Problem highlights an additional way in which conceptual reasoning can enter quantitative problem solving: *between* deciding which equations to use and evaluating the final answer. For these reasons, we urge both researchers and instructors to revisit the standard problem-solving procedures discussed above. Perhaps, instead of encouraging students (implicitly or explicitly) to treat the processing of equations as symbol manipulation, we should help students learn to spot and exploit opportunities where blended processing can help them find shortcuts and gain a deeper understanding of the physical meaning of their solutions.

**Two examples of symbolic forms as an instructional target**

Our analysis also helps us refine our vague instructional suggestion to "teach students to blend conceptual and symbolic reasoning" into something more concrete.

Symbolic forms-based reasoning, we argue, is a productive and feasible target of instruction, and techniques for helping students engage in such reasoning exist. Some researchers (such as Redish and Hammer (2009)) discuss instructional strategies intended to help students see the conceptual ideas in equations, yet they provide no specific examples. To build on this literature, we provide two concrete examples from our teaching.

The first example comes from a large lecture, undergraduate, introductory physics course. The instructors (this article's third author and another colleague[6]) use a modified treatment of a classic question in order to emphasize the symbolic forms-based reasoning implicit in the usual explanation. The question is this: A ranger aims a tranquilizer gun directly at a monkey, who is hanging from a branch. At the moment the gun is fired, the monkey drops from the branch. Will the tranquilizer dart hit the monkey?

The answer is that the dart hits the monkey, even though the monkey drops. Here is the usual reasoning. First, imagine the same scenario but with no gravitational force: The dart travels in a straight line and hits the motionless monkey. Now, consider how "turning on" gravitational force modifies this scenario. While the dart travels to the monkey, the monkey falls a certain distance; but during this time, the dart also falls by the *same* distance below that straight line — below where it would have been if there were no gravitational force. Usually, instructors use this scenario to illustrate the independence of the horizontal and vertical components of motion; the vertical distance the monkey falls equals the vertical distance the dart "falls" below where it would have been in the absence of gravity, because the dart's horizontal motion does not affect the vertical displacement to its motion caused by the gravitational force.

$$Height = \begin{bmatrix} Height \ where \ the \ object \ would \\ have \ been \ with \ no \ gravity \end{bmatrix} - [Distance \ it \ falls \ due \ to \ gravity]$$

$$= \begin{bmatrix} Height \ where \ the \ object \ would \\ have \ been \ with \ no \ gravity \end{bmatrix} - \frac{1}{2}gT^2$$

Figure 2: The equation describing the height of the dart and of the monkey.

Building on this conceptual insight, the instructor highlights the *Base + Change* symbolic form that is implicit in this reasoning. Specifically, the instructor emphasizes that the height of either the monkey or the dart at time $T$ (when the dart reaches the monkey) can be written as the equation in Fig. 2. The instructor discusses how this equation "says" that the *final* height equals the *base* height (in the absence of gravitational force) minus the *change* in height due to gravitational force. Since the *base* heights are the same for the monkey and dart (i.e., the dart would hit the monkey in the absence of gravitational force), and because the height of each object *changes* by the same amount due to the gravitational force, $-(1/2)gT^2$, the final heights are the same for each. So, the dart hits the monkey. In this way, the instructor illustrates the *Base + Change* symbolic form and shows how it affords a calculation-free shortcut highlighting the connection to the underlying physical processes in this scenario.

Our second instructional example comes from the algebra-based introductory physics course described by Redish and Hammer (2009). In the discussion sections of this course, students engage in small-group collaborative learning using "tutorials," which are guided-inquiry worksheets (Elby et al., 2007). The Momentum Tutorial leads students to figure out the formula for "oomph," which turns out to be the physics concept of momentum, using their intuitive ideas of motion. First, students consider a rock and pebble thrown at the same speed. The tutorial asks, "Which one has more oomph?" It goes on to ask if the rock is twice as massive as the pebble, *"intuitively*, how does the rock's oomph compare to the pebble's oomph?" Students generally have no trouble formulating the idea that the more massive the rock is compared to the pebble, the greater its "oomph" as compared to the pebble's, sometimes expressing their ideas qualitatively, sometimes in more mathematical terms (e.g., "proportional").

Next, the tutorial asks students to consider two identical bowling balls rolling at different speeds. If the faster ball is exactly 7 times as fast as the slower one, "intuitively, how does the faster ball's oomph compare to the slower ball's oomph?" Again, most students find it obvious that the faster ball has more oomph and that 7 times faster probably corresponds to 7 times as much oomph.

Finally, this section of the tutorial asks,

> The physics concept corresponding to oomph is momentum. Building on your above answers, figure out a formula for momentum (oomph) in terms of mass and velocity. Explain how the formula expresses your intuitions from... above.

Almost all student groups we have observed come up with the correct momentum equation, $p = mv$, and can articulate the intuitive conceptual schema underlying it, the idea that having more mass or more speed causes the object to have more oomph. We view this tutorial as helping students construct and/or use the *Prop+* (positive proportionality) symbolic form (Sherin, 2001, p. 533), which combines the "direct proportionality" symbol template with the intuitive conceptual schema of one quantity increasing if another one does through a cause-and-effect relationship.

These instructional examples echo prior research suggesting that symbolic forms-based reasoning is something that can be scaffolded in instruction. For instance, Iszak (2004) documented eighth graders using *Base + Change* to construct a mathematical equation after interacting in a rich learning environment for several hours. Almost all of Sherin's undergraduate subjects displayed evidence of using symbolic forms (Sherin, 2001, 2006). Tuminaro and Redish (2007), studying students in a physics course for undergraduate life science and non-science majors, documented the use of symbolic forms-based reasoning during problem solving. For these reasons, we believe that some students' failure to blend conceptual and symbolic reasoning during mathematical processing reflects not a lack of ability, but a lack of scaffolding. As discussed above, the standard problem-solving procedures taught in textbooks and endorsed by researchers fail to support such scaffolding.[7] Indeed, such procedures may hinder students' development and use of symbolic forms, by implicitly encouraging students to view equations merely as tools for computation and symbol manipulation.

## Conclusion

A gap exists in the literature: research on quantitative problem solving has focused on how experts and novices *select* equations but not on how they *use* the selected equations to solve problems. This paper attempts to address this gap with an illustrative case study showing how two students process the same physics equation differently.

Analyzing how Alex and Pat explained and used a standard kinematic equation, $v = v_0 + at$, we attributed part of the difference in their reasoning patterns to the use or lack of use of a knowledge element called a *symbolic form*. A symbolic form is a blend of symbolic and conceptual knowledge, a "marriage" of a symbol template to an intuitive conceptual schema. Pat's use of a symbolic form enabled him to give an intuitive explanation of the velocity equation and to quickly find a non-computational shortcut to the Two Balls Problem. Alex's explanation and problem solving, although productive and correct, were more procedural, and her processing of the velocity equation was more computational. So, as we argued, Pat's solution to the Two Balls Problem shows more expertise. However, it is Alex's solution that aligns more closely with the standard problem-solving procedures advocated by researchers and taught to students. We have used this result, along with our sample instructional techniques, to argue that blending conceptual and symbolic reasoning can be a desirable and feasible instructional target.

Given our arguments, a researcher might take away the message that we advocate replacing the "mathematical processing" step in a problem solving procedure with a "symbolic forms-based reasoning" step. This is not what we are suggesting. Not all quantitative physics questions have a non-computational shortcut as the Two Balls Problem does. However, as we argued, a valuable piece of Pat's solution to the Two Balls Problem is the adaptive expertise displayed by his evaluation of multiple solution paths. Good problem solving involves making decisions, not just following a set procedure (Reif, 2008). Aligning with this argument, we support a model of expert problem solving that does not always require either symbol manipulation or symbolic forms-based reasoning. Instead, good problem solvers have these and other tools in their toolbox, and they select which tools to use based on the details of the problem (Reif, 2008; Reif & Heller, 1982; Schoenfeld, 1985, 1992).

**REFERENCES**

Arcavi, A. (1994). Symbol sense: Informal Sense-Making in Formal Mathematics. *For the Learning of Mathematics*, *14*(3), 24–35.

Chi, M. T. H., Feltovich, P. J., & Glaser, R. (1981). Categorization and representation of physics problems by experts and novices*. *Cognitive science*, *5*(2), 121–152.

Chi, M. T. H., Glaser, R., & Rees, E. (1982). Expertise in problem solving. In R. J. Sternberg (Ed.), *Advances in the psychology of human intelligence* (Vol. 1). Hillsdale, NJ: Erlbaum.

Dhillon, A. S. (1998). Individual differences within problem-solving strategies used in physics. *Science Education*, *82*(3), 379–405.

Dufresne, R. J., Gerace, W. J., Hardiman, P. T., & Mestre, J. P. (1992). Constraining novices to perform expertlike problem analyses: Effects on schema acquisition. *The Journal of the Learning Sciences*, *2*(3), 307–331.

Elby, A., Scherr, R. E., McCaskey, T., Hodges, R., Redish, E. F., Hammer, D., & Bing, T. (2007). Open Source Tutorials in Physics Sensemaking: Suite I. Retrieved from http://www.spu.edu/depts/physics/tcp/tadevelopment.asp

Fauconnier, G., & Turner, M. (2003). *The Way We Think: Conceptual Blending and the Mind's Hidden Complexities*. New York, NY: Basic Books.

Gee, J. (1999). *An introduction to discourse analysis: Theory and method*. New York, NY: Routledge.

Giancoli, D. C. (2008). *Physics for Scientists and Engineers with Modern Physics* (4th ed.). London: Prentice Hall.

Hatano, G., & Inagaki, K. (1986). Two courses of expertise. In H. Stevenson, H. Azuma, & K. Hakuta (Eds.), *Child development and education in Japan*. (pp. 262–272). NY: Freeman.

Heller, J. I., & Reif, F. (1984). Prescribing effective human problem-solving processes: Problem description in physics. *Cognition and Instruction*, *1*(2), 177–216.

Heller, P., Keith, R., & Anderson, S. (1992). Teaching problem solving through cooperative grouping. Part 1: Group versus individual problem solving. *American Journal of Physics*, *60*(7), 627–636.

Hestenes, D. (2010). Modeling Theory for Math and Science Education. In R. Lesh, P. L. Galbraith, C. R. Haines, & A. Hurford (Eds.), *Modeling Students' Mathematical Modeling Competencies* (pp. 13–41). New York: Springer.

Hestenes, D., Wells, M., & Swackhamer, G. (1992). Force concept inventory. *The physics teacher*, *30*(3), 141–158.

Hsu, L., Brewe, E., Foster, T. M., & Harper, K. A. (2004). Resource Letter RPS-1: Research in problem solving. *American Journal of Physics*, *72*(9), 1147–1156.

Huffman, D. (1997). Effect of Explicit Problem Solving Instruction on High School Students' Problem-Solving Performance and Conceptual Understanding of Physics. *Journal of Research in Science Teaching*, *34*(6), 551–570.

Izsák, A. (2004). Students' Coordination of Knowledge When Learning to Model Physical Situations. *Cognition and Instruction*, *22*(1), 81–128.

Larkin, J. H., McDermott, J., Simon, D. P., & Simon, H. A. (1980). Expert and novice performance in solving physics problems. *Science*, *208*(4450), 1335–1342.

Larkin, J. H., & Reif, F. (1979). Understanding and teaching problem-solving in physics. *International Journal of Science Education*, *1*(2), 191–203.

Maloney, D. P. (1994). Research on problem solving: Physics. In D. . Gabel (Ed.), *Handbook of research on science teaching and learning* (pp. 327–354). New York: Macmillan.

Maloney, D. P. (2011). An Overview of Physics Education Research on Problem Solving. In Getting Started in PER (1,2). Retrieved from http://www.compadre.org/Repository/document/ServeFile.cfm?ID=11457&DocID=2427

McDermott, L. C. (1991). Millikan Lecture 1990: What we teach and what is learned—Closing the gap. *American Journal of Physics*, *59*(4), 301–315.

McDermott, L. C., & Redish, E. F. (1999). Resource letter: PER-1: Physics education research. *American Journal of Physics*, *67*(9), 755–767.

Miles, M. B., & Huberman, A. M. (1994). *Qualitative Data Analysis* (2nd ed.). London: SAGE Publications.

Mualem, R., & Eylon, B. S. (2010). Junior high school physics: Using a qualitative strategy for successful problem solving. *Journal of Research in Science Teaching*, *47*(9), 1094–1115.

Redish, E. F., & Hammer, D. (2009). Reinventing college physics for biologists: Explicating an epistemological curriculum. *American Journal of Physics*, *77*(7), 629–642.

Redish, E. F., & Smith, K. A. (2008). Looking beyond content: Skill development for engineers. *Journal of Engineering Education*, *97*(3), 295–307.

Reif, F. (2008). *Applying Cognitive Science to Education*. Cambridge, MA: MIT Press.

Reif, F., & Heller, J. I. (1982). Knowledge Structure and Problem Solving in Physics. *Educational Psychologist*, *17*(2), 102–127.

Schoenfeld, A. H. (1985). *Mathematical Problem Solving*. New York: Academic Press.

Schoenfeld, A. H. (1992). Learning to think mathematically: Problem solving, metacognition, and sense making in mathematics. In D. Grouws (Ed.), *Handbook of Research on Mathematics Teaching and Learning* (pp. 334–370). New York: MacMillan.

Sherin, B. (2001). How Students Understand Physics Equations. *Cognition and Instruction*, *19*(4), 479–541.

Sherin, B. (2006). Common sense clarified: The role of intuitive knowledge in physics problem solving. *Journal of research in science teaching*, *43*(6), 535–555.

Shin, N., Jonassen, D. H., & McGee, S. (2003). Predictors of well-structured and ill-structured problem solving in an astronomy simulation. *Journal of research in science teaching*, *40*(1), 6–33.

Simon, D. P., & Simon, H. A. (1978). Individual differences in solving physics problems. In R. S. Sigler (Ed.), *Children's thinking: What develops?* (pp. 325–348). Hillsdale, NJ: Lawrence Erlbaum.

Taasoobshirazi, G., & Glynn, S. M. (2009). College students solving chemistry problems: A theoretical model of expertise. *Journal of Research in Science Teaching*, *46*(10), 1070–1089.

Tannen, D. (1993). *Framing in discourse*. New York: Oxford University Press.

Tuminaro, J., & Redish, E. F. (2007). Elements of a cognitive model of physics problem solving: Epistemic games. *Physical Review Special Topics-Physics Education Research*, *3*(2), 020101.

Van Heuvelen, A. (1991a). Overview, Case Study Physics. *American Journal of Physics*, *59*(10), 898–907.

Van Heuvelen, A. (1991b). Learning to think like a physicist: A review of research-based instructional strategies. *American Journal of Physics*, *59*(10), 891–897.

VanLehn, K., & van de Sande, B. (2009). Acquiring Conceptual Expertise from Modeling: The Case of Elementary Physics. In K. A. Ericsson (Ed.), *The Development of Professional Performance: Toward Measurement of Expert Performance and Design of Optimal Learning Environments* (pp. 356–378). Cambridge, UK: Cambridge University Press.

Walsh, L. N., Howard, R. G., & Bowe, B. (2007). Phenomenographic Study of Students. *Physical Review Special Topics-Physics Education Research*, *3*(2), 020108.

Wertheimer, M. (1959). *Productive Thinking*. New York: Harper and Row.

Young, H. D., & Freedman, R. A. (2003). *University Physics with Modern Physics with Mastering Physics* (11th ed.). San Francisco: Addison Wesley.

**FOOTNOTES**

[1] We do not mean to imply, however, that completing these steps necessarily or even commonly involves symbolic forms-based reasoning. Using an *informal* conceptual schema to *invent* an equation differs from common instantiations of the "conceptual reasoning" step, in which students generally use *formal* concepts to *select* equations. And giving a physical interpretation of a mathematical answer often involves attaching physical significance to a number rather than attaching meaning to a functional relation between variables.

[2] Through doing the interviews, we realized that for any reasonable fourth floor balcony, the two balls would hit the ground before 5 seconds. However, this did not present a significant problem as few students mentioned this issue in their solution, and the ones that did proceeded to solve the problem as if the balls would not hit the ground.

[3] We note that Alex incorrectly labels the acceleration of the dropped ball with units of m/s instead of $m/s^2$. However, we do not focus on this mistake because it does not propagate into her manipulation of the equations, which is the focus of our analysis.

[4] It is possible that Pat's symbolic forms-based interpretation of the velocity equation was not active in his reasoning when he solved the Two Balls Problem. Indeed, several of our interviewees did not reason consistently across the two prompts. However, the coherence of Pat's reasoning across the two prompts suggests, though does not prove, that he was interpreting the velocity in the same way in both segments of the interview.

[5] Although our sample size was not large enough to do a correlational study, readers may be interested to know how many students, like Pat, both answered the Two Balls Problem with blended processing  and explained the velocity equation with symbolic forms-based reasoning; and likewise, how many students answered both questions as Alex did.  Of the 10 other unambiguously coded students, three students solved the Two Balls Problem with blended processing.  All three of these students also gave a symbolic forms-based explanation of the velocity equation to a friend.  Of the seven students who did not solve the Two Balls Problem through blended processing, five did not give a symbolic forms-based explanation of the velocity equation, while two did. Future work might sample larger populations to generalize a connection between a *Base + Change* symbolic forms-based explanation of the velocity equation and blended processing on the Two Balls Problem.

[6] This instructional modification is courtesy of David Hammer, the other instructor for the course.

[7] While research has shown that students exhibit symbolic forms-based reasoning in ways consistent with existing problem solving schemes, such schemes do not explicitly support symbolic forms-based reasoning.  In translating a physical scenario into equations, students are encouraged to use formal physics concepts to select from equations they already know, not to use intuitive schema to invent or interpret equations. While productive, this kind of translation sidesteps the blending of conceptual and formal

mathematical reasoning. And in checking their mathematical answers, students are encouraged to check if a numerical answer is physically plausible, which taps into intuitive knowledge but not into intuitive conceptual schemata. To be fair, however, students are also sometimes encouraged to check the functional relations in their final symbolic expression for plausibility, which can definitely involve the use of symbolic forms corresponding to direct and inverse proportionality. We advocate engaging students in this kind of answer checking more often.